\input harvmac
\input amssym

\def\mod{{\rm mod}}


\def\IL{\relax{\rm I\kern-.18em L}}
\def\IH{\relax{\rm I\kern-.18em H}}
\def\IR{\relax{\rm I\kern-.18em R}}
\def\IC{\relax\hbox{$\inbar\kern-.3em{\rm C}$}}
\def\IZ{\relax\ifmmode\mathchoice
{\hbox{\cmss Z\kern-.4em Z}}{\hbox{\cmss Z\kern-.4em Z}}
{\lower.9pt\hbox{\cmsss Z\kern-.4em Z}} {\lower1.2pt\hbox{\cmsss
Z\kern-.4em Z}}\else{\cmss Z\kern-.4em Z}\fi}

\def\CN {{\cal N}}

\def\CJ {{\cal J}}

\def\CE {{\cal E}}


\def\CN {{\cal N}}

\def\CE{{\cal E }}

\font\manual=manfnt \def\dbend{\lower3.5pt\hbox{\manual\char127}}

\def\IZ{\relax\ifmmode\mathchoice
{\hbox{\cmss Z\kern-.4em Z}}{\hbox{\cmss Z\kern-.4em Z}}
{\lower.9pt\hbox{\cmsss Z\kern-.4em Z}} {\lower1.2pt\hbox{\cmsss
Z\kern-.4em Z}}\else{\cmss Z\kern-.4em Z}\fi}
\def\half {{1\over 2}}

\def\bar{\overline}

\def\rt2{\sqrt{2}}
\def\irt2{{1\over\sqrt{2}}}

\def\hat{\widehat}
\def\slashchar#1{\setbox0=\hbox{$#1$}           
   \dimen0=\wd0                                 
   \setbox1=\hbox{/} \dimen1=\wd1               
   \ifdim\dimen0>\dimen1                        
      \rlap{\hbox to \dimen0{\hfil/\hfil}}      
      #1                                        
   \else                                        
      \rlap{\hbox to \dimen1{\hfil$#1$\hfil}}   
      /                                         
   \fi}
\lref\bertolini{
  M.~Bertolini and M.~Trigiante,
  ``Microscopic entropy of the most general four-dimensional BPS black  hole,''
  JHEP {\bf 0010}, 002 (2000)
  [arXiv:hep-th/0008201].
} \lref\ddmp{G.~Lopes Cardoso, B.~de Wit and T.~Mohaupt,
  ``Corrections to macroscopic supersymmetric black-hole entropy,''
  Phys.\ Lett.\ B {\bf 451}, 309 (1999)
  [arXiv:hep-th/9812082];
H.~Ooguri, A.~Strominger and C.~Vafa,
  ``Black hole attractors and the topological string,''
  Phys.\ Rev.\ D {\bf 70}, 106007 (2004)
  [arXiv:hep-th/0405146];

  A.~Dabholkar, F.~Denef, G.~W.~Moore and B.~Pioline,
  ``Exact and asymptotic degeneracies of small black holes,''
  arXiv:hep-th/0502157;

  G.~L.~Cardoso, B.~de Wit, J.~Kappeli and T.~Mohaupt,
  ``Asymptotic degeneracy of dyonic N = 4 string states and black hole
  entropy,''
  JHEP {\bf 0412}, 075 (2004)
  [arXiv:hep-th/0412287];
A.~Dabholkar,
  ``Exact counting of black hole microstates,''
  arXiv:hep-th/0409148;

  E.~Verlinde,
  ``Attractors and the holomorphic anomaly,''
  arXiv:hep-th/0412139;
}

\lref\DVV{
  R.~Dijkgraaf, E.~Verlinde and H.~Verlinde,
  ``Counting dyons in N = 4 string theory,''
  Nucl.\ Phys.\ B {\bf 484}, 543 (1997)
  [arXiv:hep-th/9607026].
}

\lref\finn{
  V.~Balasubramanian and F.~Larsen,
  ``On D-Branes and Black Holes in Four Dimensions,''
  Nucl.\ Phys.\ B {\bf 478}, 199 (1996)
  [arXiv:hep-th/9604189].
}

\lref\vijay{
  V.~Balasubramanian,
  ``How to count the states of extremal black holes in N = 8 supergravity,''
  arXiv:hep-th/9712215.
}

\lref\msw{
  J.~M.~Maldacena, A.~Strominger and E.~Witten,
  ``Black hole entropy in M-theory,''
  JHEP {\bf 9712}, 002 (1997)
  [arXiv:hep-th/9711053].
} \lref\GaiottoGF{
  D.~Gaiotto, A.~Strominger and X.~Yin,
  ``New connections between 4D and 5D black holes,''
  arXiv:hep-th/0503217.
}\lref\ssy{
  D.~Shih, A.~Strominger and X.~Yin,
  ``Recounting dyons in N = 4 string theory,''
  arXiv:hep-th/0505094.
} \lref\kallk{
  R.~Kallosh and B.~Kol,
  ``E(7) Symmetric Area of the Black Hole Horizon,''
  Phys.\ Rev.\ D {\bf 53}, 5344 (1996)
  [arXiv:hep-th/9602014].
} \lref\CremmerUP{
  E.~Cremmer and B.~Julia,
  ``The SO(8) Supergravity,''
  Nucl.\ Phys.\ B {\bf 159}, 141 (1979).
}

\lref\GaiottoXF{
  D.~Gaiotto, A.~Strominger and X.~Yin,
  ``New connections between 4D and 5D black holes,''
  arXiv:hep-th/0503217.
}
\lref\mms{
  J.~M.~Maldacena, G.~W.~Moore and A.~Strominger,
  ``Counting BPS black holes in toroidal type II string theory,''
  arXiv:hep-th/9903163.
}

\lref\gv{
  R.~Gopakumar and C.~Vafa,
  ``M-theory and topological strings. II,''
  arXiv:hep-th/9812127.
}
\lref\DijkgraafXW{
  R.~Dijkgraaf, G.~W.~Moore, E.~Verlinde and H.~Verlinde,
  ``Elliptic genera of symmetric products and second quantized strings,''
  Commun.\ Math.\ Phys.\  {\bf 185}, 197 (1997)
  [arXiv:hep-th/9608096].
} \lref\borch{R. E. Borcherds, "Automorphic forms on
$O_{s+2,2}(R)$ and infinite products" Invent. Math. {\bf 120}
(1995) 161.}

 \lref\KAWAI{
  T.~Kawai,
  ``$N=2$ heterotic string threshold correction, $K3$ surface and generalized
  Kac-Moody superalgebra,''
  Phys.\ Lett.\ B {\bf 372}, 59 (1996)
  [arXiv:hep-th/9512046].
}

\lref\DVV{
  R.~Dijkgraaf, E.~Verlinde and H.~Verlinde,
  ``Counting dyons in N = 4 string theory,''
  Nucl.\ Phys.\ B {\bf 484}, 543 (1997)
  [arXiv:hep-th/9607026].
}

\lref\gava{
  I.~Antoniadis, E.~Gava, K.~S.~Narain and T.~R.~Taylor,
  Nucl.\ Phys.\ B {\bf 455}, 109 (1995)
  [arXiv:hep-th/9507115].
}

\lref\fk{
  S.~Ferrara and R.~Kallosh,
  ``Universality of Supersymmetric Attractors,''
  Phys.\ Rev.\ D {\bf 54}, 1525 (1996)
  [arXiv:hep-th/9603090].
}

\lref\BMPV{
  J.~C.~Breckenridge, R.~C.~Myers, A.~W.~Peet and C.~Vafa,
  ``D-branes and spinning black holes,''
  Phys.\ Lett.\ B {\bf 391}, 93 (1997)
  [arXiv:hep-th/9602065].
}

\lref\ascv{ A.~Strominger and C.~Vafa,
  ``Microscopic Origin of the Bekenstein-Hawking Entropy,''
  Phys.\ Lett.\ B {\bf 379}, 99 (1996)
  [arXiv:hep-th/9601029].}

\newbox\tmpbox\setbox\tmpbox\hbox{\abstractfont }
\noblackbox
 \Title{\vbox{\baselineskip12pt\hbox to\wd\tmpbox{\hss
}}\hbox{hep-th/0506151 }} {\vbox{\centerline{Counting Dyons in
$\CN=8$ String Theory}\medskip\centerline{} }}

\centerline{David Shih,\footnote{*}{Permanent address: Department
of Physics, Princeton University, Princeton, NJ 08544, USA.}~
Andrew Strominger\footnote{**}{Permanent address: Jefferson
Physical Laboratory, Harvard University, Cambridge, MA 02138,
USA.} and Xi Yin** }
\smallskip\centerline{Center of Mathematical Sciences}
\centerline{ Zhejiang University, Hangzhou 310027 China}

\centerline{} \vskip.4in \centerline{\bf Abstract} {A recently
discovered relation between 4D and 5D black holes is used to
derive exact (weighted) BPS black hole degeneracies for  4D
$\CN=8$ string theory from the exactly known 5D degeneracies. A
direct 4D microscopic derivation in terms of weighted 4D D-brane
bound state degeneracies is sketched and found to agree.

 } \vskip.2in

\Date{} \vfill  \listtoc \writetoc
\newsec{Introduction}

In this paper, we deduce an exact formula for the modified
elliptic genus of string theory in four dimensions with $\CN=8$
supersymmetry. The modified elliptic genus, as we review below,
provides a weighted count of BPS states of $\CN=8$ string theory.
We derive a formula for it using a recently proposed exact
relation \GaiottoGF\ between 4D and 5D BPS degeneracies, together
with the known degeneracies \mms\ in 5D. In addition we sketch a
direct microscopic counting of D0-D2-D4 bound states which gives
the same result. Our hope is that this example will provide a
useful laboratory for testing the string theory relations recently
proposed in e.g.\ddmp.

Some years ago an explicit formula for the elliptic genus for BPS
states in 4D $\CN=4$ theories was presciently conjectured \DVV.
This formula was recently derived using the 4D-5D connection in
\ssy. The present work is an extension of \ssy\ to 4D $\CN=8$
theories. Previous work in this direction includes
\refs{\finn,\vijay,\bertolini}.

In the next section we review the 5D index defined and computed in
\mms. In section 3 we use the 4D-5D connection to derive the 4D
index. In section 4 we sketch how this expression should follow
(for one element of the U-duality class of black holes) from a
microscopic analysis.

\newsec{Review of the 5D modified eliptic genus}

In this section, we want to summarize the work of reference \mms\
on counting the microstates of 1/8 BPS black holes in five
dimensions. These can be realized in string theory as the usual
D1-D5-momentum system of type IIB on $T^4\times S^1$, with $Q_1$
D1-branes, $Q_5$ D5-branes and integral $S^1$ momentum $n$. The
reason that microstate counting of this system is more difficult
than for $K3$ compactification is because the usual supersymmetric
index that counts these microstates, the orbifold elliptic genus
of $Hilb^k(K3)$ with $k=Q_1Q_5$, vanishes when $K3$ is replaced
with $T^4$. In \mms, this difficulty was overcome by defining (and
then computing) a new supersymmetric index $\CE_2$, closely
related with the elliptic genus, which is nonvanishing for $T^4$.
We will refer to this new supersymmetric index as the modified
elliptic genus of $Hilb^k(T^4)$. It is defined to be
\eqn\CEtwodef{ \CE_2^{(k)} =
Tr\left[(-1)^{2J_L^3-2J_R^3}2(J_R^3)^2q^{L_0}\bar q^{\bar
L_0}y^{2J_L^3}\right] } where the trace is over states of the
sigma model with target space $Hilb^k(T^4)$.\foot{A free sigma
model on $R^4\times T^4$ is factored out here, and our definition
differs by a factor of 2 from \mms.} Here $J_L^3$ and $J_R^3$ are
the left and right half-integral $U(1)$ charges of the CFT, and
they are identified with generators of SO(4) rotations of the
transverse $R^4$. The $S^1$ momentum is $n=L_0-\bar L_0$.  The
usual elliptic genus is given by the same formula but without the
$2(J_R^3)^2$ factor; it is these two insertions of $J_R^3$ that
make $\CE_2$ nonvanishing for $T^4$.

As for $K3$, here it is convenient to define a generating function
for the modified elliptic genus: \eqn\generating{ \CE_2 =
\sum_{k\ge 1} p^k \CE_2^{(k)} } In \mms, this was shown to be
given by the following sum \eqn\generatingsum{  \CE_2(p,q,y)=
\sum_{s,k,n,\ell}s(p^kq^ny^\ell)^s\hat c(nk,\ell) } with the sum
running over $s,k\ge 1$, $n\ge0$, $\ell\in \Bbb Z$. Note that the
$\bar q$ dependence has dropped out -- only the $\bar L_0=0$
states contribute to the modified elliptic genus. Of course, the
index must have this property in order to count BPS states, since
the BPS condition is equivalent to requiring $\bar L_0=0$.

It was furthermore shown in \mms\ that the integers $\hat
c(nm,\ell)$ are the coefficients in the following Fourier
expansion \eqn\fourierexpansion{ Z(q,y)\equiv
-\eta(q)^{-6}\vartheta_1(y|q)^2 = \sum_{n,\ell}\hat
c(n,\ell)q^ny^\ell } where $\eta(q)$ is the usual Dedekind eta
function, and $\vartheta_1(y|q)$ is defined by the product formula
\eqn\deftheta{ \vartheta_1(y|q) =
i(y^{1/2}-y^{-1/2})q^{1/8}\prod_{n=1}^{\infty}(1-q^n)(1-yq^n)(1-y^{-1}q^{n})
}
Finally, it was observed in \mms\ that $\hat c(n,\ell)$ actually
only depends on a single combination of parameters $4n-\ell^2$:
\eqn\csimp{ \hat c(n,\ell) = \hat c(4n-\ell^2) } Using \csimp\ in
\generatingsum\ yields \eqn\generatingsumii{ \CE_2(p,q,y)=
\sum_{s,k,n,\ell}s(p^kq^ny^\ell)^s\hat c(4nk-\ell^2) } When
$(k,n,\ell)$ are coprime, $\hat c(4nk-\ell^2)$ counts BPS black
holes with $k=Q_1Q_5$, $S^1$ momentum $n$ and spin $J^3_L={\ell
\over 2}$, multiplied by an overall $(-)^\ell$ and summed over
$J^3_R$ weighted by $2(J_R^3)^2 (-)^{2J^3_R}$: \eqn\dsz{\hat
c(4nk-\ell^2)\Big|_{(k,n,\ell)\, {\rm coprime}} =(-)^\ell
\sum_{J_R, BPS~states}2(J_R^3)^2 (-)^{2J^3_R}}
 When they are not coprime,
the black hole can fragment, and the situation is more complicated
due to multiple contributions in $\CE_2$ \mms. In this paper we
will always avoid this complication by choosing coprime charges.

We should note that $Z(q,y)$ is also the modified elliptic genus
of $T^4$, i.e.\
\eqn\rxa{\CE_2^{(1)}=\sum_{n,\ell}\hat
c(n,\ell)q^ny^\ell=Z(q,y).} This corresponds to the coprime D1-D5
system with $k=1=Q_1=Q_5$. By writing \eqn\cenn{ Z(q,y)= \sum_{m}
\hat c(4m) q^{m}\sum_k q^{k^2} y^{2k} + \sum_{m} \hat c(4m-1)
q^{m}\sum_k q^{k^2+k} y^{2k+1} } and using \fourierexpansion\
along with the standard Fourier expansion of the theta function
\eqn\thetaexpand{
\vartheta_1(y|q) = i\sum_{n\in\Bbb
Z}(-1)^nq^{(n-1/2)^2/2}y^{n-1/2}
}
one can reorganize the generating functions for $\hat c$ as
\eqn\ccs{ \eqalign{ &\sum_m \hat c(4m) q^m
= -q^{1\over4} \eta(q)^{-6} \sum_{m\in{\Bbb Z}} q^{m^2+m}, \cr
&\sum_m \hat c(4m-1) q^m
=q^{1\over 4}\eta(q)^{-6} \sum_{m\in{\Bbb Z}} q^{m^2}.
 }}
These expressions will analyzed microscopically below in section
4.

\newsec{The 4D modified elliptic genus}

In this section we use the conjecture of \GaiottoGF\  to transform
the 5D degeneracies into 4D ones. The fact that the $\hat c$
coefficients depend only on the combination $4nk-\ell^2$ is very
encouraging, for the following reason. We expect the 1/8 BPS 5D
degeneracies to be related to degeneracies of 1/8 BPS black holes
in 4D, and in 4D U-duality implies \kallk\ that the black hole
entropy must depend on the unique quartic invariant of $E_{7,7}$,
the so-called Cremmer-Julia invariant \CremmerUP. In an $\CN=4$
language, this invariant takes the form \eqn\cjinv{ \CJ
 = q_e^2q_m^2-(q_e\cdot
q_m)^2 } where $q_e$ and $q_m$ are the electric and magnetic
charge vectors for $\CN=4$ BPS states. (See e.g.\ \ssy\ for
details on the notation.) This is precisely the dependence of
$\hat c$ on $n$, $m$, $\ell$, provided we identify
\eqn\nmlidentify{ k = {1\over2}q_e^2,\qquad n =
{1\over2}q_m^2,\qquad \ell = q_e\cdot q_m .}  Note that from the
purely 5D point of view, there was no obvious reason that $\hat c$
should depend only on the combination $4nk-\ell^2$ as there is no
5D U-duality which mixes spins with charges.

Let us now derive the identification \nmlidentify\ from the
dictionary of \GaiottoGF, beginning from the IIB spinning 5D
D1-D5-n black hole of the previous section. First we T-dual on
$S^1$ to obtain a black hole with spin $\ell \over 2$, F-string
winding $n$, $Q_1$ D0-branes, and $Q_5$ D4-branes. Now T-dual so
that there are $Q_1+Q_5$ D2 branes with intersection number
$Q_1Q_5=k$ on the $T^4$. Next we compactify on a single center
Taub-NUT, whose asymptotic circle we identify as the the new
M-theory circle. The result is three orthogonal sets of $(n,
Q_1,Q_5)$ D2-branes on $T^6$, $\ell$ D0-branes, and one D6-brane.
For IIA D-brane configurations with D0, D2, D4, D6 charges
$(q_0,q_{ij},p^{ij},p^0)$, where $i=1,...6$ runs over the $T^6$
cycle and $p^{ij}=-p^{ji},~q_{ij}=-q_{ji}$ $\CJ$ reduces
to\foot{See e.g. \fk, equation (66), and take
$p^0=p_{87},~~p_{8i}=0$, etc. Our definition of $\CJ$ differs from
that of \fk\ by a sign. } \eqn\tfc{\eqalign{\CJ={1\over 12}&(
q_0\epsilon_{ijklmn}p^{ij}p^{kl}p^{mn}+p^0\epsilon^{ijklmn}q_{ij}q_{kl}q_{mn})\cr
& -p^{ij}q_{jk}p^{kl}q_{li}+{1 \over 4}p^{ij}q_{ij}p^{kl}q_{kl}-
(p^0q_0)^2+{1 \over 2}p^0q_0p^{ij}q_{ij}.\cr}} For our D0-D2-D6
configuration, we can pick a basis of cycles without loss of
generality such that the nonzero charges are
\eqn\Dnot{
p^0=0,\quad q_0=\ell,\quad q_{12}=-q_{21}=n,\quad
q_{34}=-q_{43}=Q_1,\quad q_{56}=-q_{65}=Q_5
}
Then \tfc\ reduces to \eqn\fgi{ \CJ =4nk-\ell^2,} which, as stated
above, is exactly the argument of \generatingsum.

According to \GaiottoGF\ the weighted degeneracy of the 4D black
hole resulting from U-duality and Taub-NUT compactification equals
that of the original 5D black hole, when $J_R^3$ in \dsz\ is
identified with the generator $J^3$ of ${\Bbb R}^3$ rotations in
4D. Note that, since $\CJ$ is odd if and only if $\ell$ is, we may
trade $(-)^\ell$ for $(-)^\CJ$ in \dsz.  Therefore, for fixed
coprime charges, the weighted 4D BPS degeneracy depends only on
the the Cremmer-Julia invariant and is given by\eqn\dszd{
 \sum_{J^3, BPS~states}2(J^3)^2 (-)^{2J^3}=(-)^\CJ\hat c(\CJ).}
Note that, although this formula for the 4D BPS degeneracy was
derived assuming a specific D6-D2-D0 configuration, it applies to
all D-brane configurations by U-duality.

   As a first check on this conjecture, we note that for large charges $\hat c (\CJ)
   \sim e^{\pi \sqrt{J}}$. From the supergravity solutions ${\rm
   Area}=4\pi \sqrt{J}$, so there is agreement with the
   Bekenstein-Hawking entropy.

As an example, let's consider the modified elliptic genus for the
D4-D0 black hole on $T^6$, in which we fix the D4 charges and sum
over D0 charge $q_0$. Consider the $T^6$ of the form $T^2\times
T^2\times T^2$ with $\alpha_1, \alpha_2,\alpha_3$ being the three
2-cycles associated with the $T^2$'s. Let $A^1,A^2,A^3$ be the
dual 4-cycles. We shall consider the D4-brane wrapped on the cycle
$[P]=A^1+A^2+A^3$. Its triple self-intersection number is
$D=P\cdot P\cdot P=6$. From \tfc\ we have \eqn\ffk{\CJ={4q_0 }.}
We then have \eqn\genfn{\CE_2(q) =\sum_{q_0\in\Bbb Z}\hat c
(4q_0)q^{q_0}=-q^{1\over4} \eta(q)^{-6} \sum_{m\in{\Bbb Z}}
q^{m^2+m}.} according to \ccs.

A straightforward generalization of this example is the D4-D2-D0
system, where we wrap $(q_1,q_2,q_3)$ D2 branes on the 2-cycles
$(\alpha_1,\alpha_2,\alpha_3)$. In this case, the Cremmer-Julia
invariant becomes
\eqn\dtwocj{
\CJ =4(q_0+q_1q_2+q_1q_3+q_2q_3)-(q_1+q_2+q_3)^2
}
and the sum over $q_0$ produces
\eqn\sumqdtwo{
\CE_2(q) = \sum_{q_0\in \Bbb Z}(-1)^{\CJ}\hat c(\CJ)q^{q_0} =
\cases{-q^{1\over4} \eta(q)^{-6} \sum_{m\in{\Bbb Z}}
q^{m^2+m-{1\over4}\tilde\CJ} &\quad $q_1+q_2+q_3$ even\cr
 -q^{1\over 4}\eta(q)^{-6} \sum_{m\in{\Bbb Z}}q^{m^2-{1\over4}\tilde\CJ-{1\over4}} &\quad $q_1+q_2+q_3$ odd\cr}
}
where $\tilde\CJ  = 4(q_1q_2+q_1q_3+q_2q_3)-(q_1+q_2+q_3)^2$. Now
let us turn to the 4D derivation of \genfn\ and \sumqdtwo.

\newsec{Microscopic derivation in 4D}

In this section we sketch a derivation of \genfn\ and \sumqdtwo\
using a 4D microscopic analysis. The derivation is not complete
because, as we will discuss below,  we ignore some potential
subtleties associated to the fact that $P$ is not simply
connected. In principle it should be possible to close this gap. A
microscopic description of $T^6$ black holes using the M-theory
picture of wrapped fivebranes has been given in \bertolini ,
adapting the description given in \msw\ for a general Calabi-Yau,
in terms of a $(0,4)$ 2D CFT living on the M-theory circle. For
uniformity and simplicity of presentation, we here will use the
IIA description in which fivebrane momenta around the M-theory
circle become bound states of D0 branes to D4 branes.

As above \ffk\ we examine the special case of the D4-D0 system
wrapped on $[P]=A^1+A^2+A^3$.
 The D4-D0 system can
be described in terms of the quantum mechanics of $q_0$ D0-branes
living on the D4-brane world volume $P$. The D4-brane world volume
$P$ is holomorphically embedded in the $T^6$. One can compute its
Euler character, $\chi(P)=6$. It follows from the Riemann-Roch
formula that the only modulus of $P$ is the overall translation in
$T^6$.\foot{ The dual line bundle ${\cal L}_P$ of the divisor $P$
has only one holomorphic section. However as $T^6$ is not simply
connected, the line bundle ${\cal L}_P$ is not only determined by
$c_1({\cal L}_P)=[P]$. In fact the translation of $T^6$ takes it
to a different line bundle.  } Since $\chi(P)=6$, $P$ has $4+2b_1$
2-cycles. By the Lefschetz hyperplane theorem we have
$b_1(P)=b_1(T^6)=6$, and therefore $b_2(P)=16$. All but one of the
2-cycles come from the intersection of $P$ with ${6\choose 4}=15$
4-cycles in $T^6$. We will be mostly interested in 3 of these,
denoted by $\tilde\alpha_i$, corresponding to intersections of
$A^i$ with $P$. Turning on fluxes along these three 2-cycles
corresponds to having charges of D2-branes wrapped on the
$\alpha_i$'s. Their intersection numbers are \eqn\intera{ \tilde
\alpha_i\cdot \tilde\alpha_j = \left\{ \eqalign{ 0,~~~i=j\cr
1,~~~i\not=j } \right. } There is, however, one extra 2-cycle in
$P$, which we shall denote by $\beta$, that does not correspond to
any cycle in the $T^6$.

One can show from the adjunction formula that $c_1(P)$ is
Poincar\'e dual to $-(\tilde\alpha_1+\tilde\alpha_2+\tilde
\alpha_3)$. It then follows from Hirzebruch signature theorem that
\eqn\hirz{ \sigma(P) = -{2\over 3}\chi(P) + {1\over 3}\int_P c_1^2
= -2. } We conclude that the intersection form on $P$ is odd (and
that $P$ is not a spin manifold). Essentially the unique way to
extend \intera\ to an odd rank 4 unimodular quadratic form is to
have an extra 2-cycle $\gamma$ with \eqn\gammainter{ \gamma\cdot
\tilde\alpha_i = 1,~~~\gamma\cdot\gamma=1. } Now if we choose
$\beta = 2\gamma-\sum\tilde\alpha_i$, we have \eqn\betainter{
\beta\cdot \tilde \alpha_i=0, ~~~ \beta\cdot \beta = -2. } Note
that $(\tilde \alpha_i,\beta)$ is not an integral basis for
$H_2(P,{\Bbb Z})$, yet $\beta$ is the smallest 2-cycle that
doesn't intersect $\tilde \alpha_i$. The total intersection form
on $P$ is the sum of this rank 4 form together with 6 copies of
$\sigma_1$ coming from the 12 other 2-cycles in $P$.

Now one can turn on gauge field flux on the D4-brane world volume
along $\beta$, which does not correspond to any D2-brane charge.
This flux nevertheless induces D0-brane charge. There is a
subtlety in the quantization of this flux. As well known, the
curvature of the D4-brane world volume induces an anomalous
D0-brane charge $-\chi(P)/24=-{1\over 4}$.   In order that the
total D0 charge be integral the flux along the cycle $\beta$ on
the D4-brane must be half-integer, i.e. of the form
$(m+\half)\beta$.
 The total induced D0-brane charge is $\Delta
q_0 = -\half (m+\half)^2 \beta\cdot \beta-{1\over 4} = m^2+m$,
which is indeed an integer.\foot{In the M-theory picture the
anomalous D0 charge is the left-moving zero point energy $-{c_L
\over 24}=-{1\over 4}$, and the 2-cycle fluxes correspond to
momentum zero modes of scalars on a Narain lattice.}

   We ignore here the facts arising form nonzero $b_1(P)$
   that there is a moduli space of flat
   connections as well as overall $T^6$ translations which must be
   quantized. These factors are treated in the language of the 2D CFT in
   \bertolini. They
   are found to lead to extra degrees of freedom
   which are however eliminated by extra gauge constraints.
   A complete microscopic derivation, not given here, would have
   to show that a careful accounting of these factors give a trivial correction to our
   result.

It is now straightforward to reproduce \genfn. Each D0-D4 bound
state is in a hypermultiplet which contributes minus one to
$Tr\big[2(J^3)^2(-)^{2J^3}\big]$. Counting the number of ways of
distributing $n$ D0-branes among the $\chi(P)=6$ ground states of
the supersymmetric quantum mechanics, and then summing over $n$,
gives the factor of $q^{1/4}\eta(q)^{-6}=\prod_{k=1}^\infty
(1-q^k)^{-6}$ in \genfn. Including finally the sum over fluxes on
$\beta$, we precisely reproduce the degeneracy \genfn!

Let us now consider the more general case of D4-D2-D0 system.
Again we shall assume $(p^1,p^2,p^3)=(1,1,1)$. The D2-brane
charges are labelled by $(q_1,q_2,q_3)$. The bound state is
described by the D4-brane with D2-brane dissolved in its world
volume. We end up with the gauge flux \eqn\gaugflux{ F =
(m+1/2)\beta + \sum_{i=1}^3 q_i \delta_i,~~~~\delta_i\cdot \tilde
\alpha_j=\delta_{ij}. } In above expression $\delta_i$ is defined
up to a shift of an integer multiple of $\beta$. Since we are
summing over $m$, this ambiguity is irrelevant. We can choose
$\delta_i=\gamma-\tilde\alpha_i$. The total induced D0-brane
charge is then \eqn\induceda{ \eqalign{ \Delta q_0 &= -\int \half
F^2 -{1\over 4} \cr &= (m+1/2)^2 + (m+1/2)\sum q_i +{1\over 2}\sum
q_i^2 - {1\over 4}\cr &= \left(m+\half + \half \sum q_i\right)^2 -
{1\over 12} D^{AB}q_Aq_B-{1\over 4}, } } where $D^{AB}$ is the
inverse matrix of $D_{AB}\equiv D_{ABC}p^C$, \eqn\dab{
D^{AB}q_Aq_B = 3(2q_1q_2+2q_2q_3+2q_3q_1-q_1^2-q_2^2-q_3^2). }
Note that ${1\over 3}D^{AB}q_Aq_B=0~ \mod~ 4$ if $\sum q_i$ is
even, and ${1\over 3}D^{AB}q_Aq_B=-1~ \mod~ 4$ if $\sum q_i$ is
odd. Therefore $\Delta q_0$ is always an integer, as expected. The
Cremmer-Julia invariant is in this case \eqn\acj{ {\cal J} =
4\left(q_0+{1\over 12}D^{AB}q_Aq_B\right). } The counting of
D0-brane states as before gives the generating function
\eqn\gencont{ \sum_{q_0} (-)^{\cal J} c({\cal J})
q^{q_0}=-\prod_{k=1}^\infty (1-q^k)^{-6} \sum_{m\in {\Bbb Z}}
q^{m^2+m-{1\over 12}D^{AB}q_Aq_B} } in the case $\sum q_i\in
2{\Bbb Z}$ and ${\cal J}\equiv0 ~\mod ~4$, and \eqn\gencb{
\sum_{q_0} (-)^{\cal J} c({\cal J}) q^{q_0}=-\prod_{k=1}^\infty
(1-q^k)^{-6} \sum_{m\in {\Bbb Z}} q^{m^2-{1\over
12}D^{AB}q_Aq_B-{1\over 4}} } in the case $\sum q_i \in 2{\Bbb
Z}+1$ and ${\cal J}\equiv-1~\mod~4$. These are precisely the
degeneracies \sumqdtwo\ we derived from 5D earlier!

\bigskip\bigskip\bigskip

\centerline{\bf Acknowledgments} This work was supported in part
by DOE grant DEFG02-91ER-40654.  We are grateful to Greg Moore and
Boris Pioline for useful correspondence.

\listrefs

\end